\documentclass[preprint,showpacs,preprintnumbers,groupedaddress,amsmath,amssymb,aps,pra]{revtex4}

\usepackage{graphicx}
\usepackage{bm}

\begin{document}

\title{Pulse Normalisation in Slow-Light Media}

\author{Bruno Macke} 
\author{Bernard S\'{e}gard} 
\email{bernard.segard@univ-lille1.fr}
\affiliation{Laboratoire de Physique des Lasers, Atomes et Mol\'{e}cules (PhLAM), 
Centre d'Etudes et de Recherches Lasers et Applications (CERLA), 
Universit\'{e} de Lille I, 59655 Villeneuve d'Ascq, France}

\date{\today}

\begin{abstract}
We analytically study the linear propagation of arbitrarily shaped light-pulses through an absorbing medium 
with a narrow transparency-window or through a resonant amplifying medium. We point out that, under certain 
general conditions, the pulse acquires a nearly Gaussian shape, irrespective of its initial shape and of 
the spectral profile of the line. We explicitly derive in this case the pulse parameters, including its 
skewness responsible for a deviation of the delay of the pulse maximum from the group delay. We illustrate 
our general results by analysing the slow-light experiments having demonstrated the largest fractional pulse-delays. 
\end{abstract}
\maketitle

\section{INTRODUCTION }

The principle underlying most of the slow-light experiments is to
exploit the steep normal dispersion of the refractive index associated
with a pronounced peak in the transmission of the medium and the correlative
reduction of the group velocity. The situation where the resulting
time-delay of the light-pulse is large compared to its duration and
can be controlled by an external laser field is of special importance
for potential applications, especially in the domain of high-speed
all-optical signal-processing. Harris and his co-workers \cite{ref1,ref2}
opened the way to such experiments by exploiting the phenomenon of
electromagnetically induced transparency (EIT) allowing one to create
a narrow transparency-window in an otherwise optically thick atomic
vapour. Using a true-shape detection of the pulses, they demonstrated
propagation velocities as slow as $c/165$ and group delays $\tau_{g}$
as long as $4.6\:\tau_{in}$ where $c$ and $\tau_{in}$ are respectively
the velocity of the light in vacuum and the full width at half-maximum
(FWHM) of the intensity-profile of the incident pulse. Much slower
velocities have been attained in subsequent EIT experiments (for reviews
see, e.g., \cite{ref3,ref4,ref5}) and in experiments involving coherent population
oscillations \cite{ref6} or other processes to induce a transparency-window
in an absorbing medium. It is however worth noticing that only few
of these experiments, all using EIT, have succeeded in giving \emph{direct}
demonstrations of fractional delays $\tau_{g}/\tau_{in}$ exceeding
unity \cite{ref2,ref7,ref8}. Theoretical discussions on the maximum time-delays
attainable in such experiments can be found in \cite{ref1,ref9,ref10}. A different
way to achieve a system with a controllable transmission-peak is to
optically induce a resonant gain in a transparent medium \cite{ref11}.
Initially proposed by Gauthier \cite{ref12}, the arrangement involving
stimulated Brillouin scattering \cite{ref13,ref14,ref15,ref16} seems particularly attractive
from the viewpoint of the above mentioned applications. The Brillouin
gain is indeed directly implemented on an optical fibre and there
are no severe constraints in the choice of the operating wavelength.
The group delay $\tau_{g}$ has already been controlled
on a range of $3.6\:\tau_{in}$ by this technique \cite{ref15}. Note that
preliminary experiments using a Raman fibre amplifier have also been
achieved \cite{ref17}. 

The purpose of our paper is to provide analytical results on the propagation
of arbitrarily shaped pulses, the central frequency of which coincides with
that of a pronounced maximum in the medium-transmission. We examine
more specifically the case where the resulting time-delays of the
pulses are large compared to their duration. Our study applies in
particular but not exclusively to the above mentioned systems. Our
approach follows in part and extends that of Bukhman \cite{ref18} with
a special attention paid to the connection of the theoretical results
with the experiments.

\section{GENERAL ANALYSIS }

We denote by $e_{in}(t)$ and $e_{out}(t)$ the slowly-varying envelopes
of the incident and transmitted pulses and by $E_{in}(\Omega)=\int_{-\infty}^{\infty}e_{in}(t)\exp(-i\Omega t)dt$
and $E_{out}(\Omega)$ their Fourier transforms. The slow-light medium
is characterised by its impulse response $h(t)$ or by its transfer
function $H(\Omega)$, Fourier transform of $h(t)$. The input/output
relation or transfer equation reads $e_{out}(t)=h(t)\otimes e_{in}(t)$
in the time-domain or $E_{out}(\Omega)=H(\Omega)E_{in}(\Omega)$ in
the frequency-domain \cite{ref19}. We assume that the incident pulse has
a finite energy, that it is not chirped ($e_{in}(t)$ real and positive)
and that $h(t)$ is also real. The local response of the medium is
characterised by the complex gain-factor $\Gamma(\Omega)=\ln\left[H(\Omega)\right]$
whose real part $F(\Omega)$ and imaginary part $\Phi(\Omega)$ are
respectively the logarithm of the medium amplitude-gain $\left|H(\Omega)\right|$
and the induced phase shift. The condition imposed to $h(t)$ implies
that $H(-\Omega)=H^{*}(\Omega)$ and thus that $F(\Omega)$ and $\Phi(\Omega)$
are respectively even and odd functions of $\Omega$. This has the
advantage of eliminating the lowest-order pulse-distortions resulting
from the gain-slope and from the group velocity dispersion at the
frequency $\omega_{0}$ of the optical carrier ($\Omega=0$). Moreover
the medium is then entirely characterised by the single real function
$h(t)$. In order to have simple expressions we use for $e_{in}(t)$
a time origin located at the pulse centre-of-gravity and for $e_{out}(t)$
a time origin retarded by the transit time at the group velocity outside
the frequency-domain of high-dispersion (local time picture). The
time delays considered hereafter are thus only those originating in
the high-dispersion region.

General properties of the transmitted pulse can be derived by Fourier
analysis. Let $x(t)$ be any of the real functions $e_{in}(t)$, $h(t)$
or $e_{out}(t)$ and $X(\Omega)$ its Fourier transform. We remark
that $X(0)=\int_{-\infty}^{\infty}x(t)dt$ and, following an usual
procedure in probability theory \cite{ref20}, we characterise $X(\Omega)$
by its cumulants $\kappa_{n}$, such that 
\begin{equation}
X(\Omega)=X(0)\exp\left(\sum_{n=1}^{\infty}\frac{\kappa_{n}}{n!}(-i\Omega)^{n}\right).\label{EQ1}
\end{equation}
For $H(\Omega)$, we see that the cumulants are simply related to
the coefficients of the series expansion of $\Gamma(\Omega)$ in powers
of $-i\Omega$ and, in particular, that $\kappa_{1}$ coincides with
the group delay $\tau_{g}=-\frac{d\Phi}{d\Omega}\mid_{\Omega=0}$.
We incidentally recall that, due to the causality principle, $\tau_{g}$
can be related to the gain profile \cite{ref21,ref22}. Within our assumptions,
this relation reads 
\begin{equation}
\tau_{g}=P\int_{-\infty}^{\infty}\frac{\ln H_{0}-\ln\left|H(\Omega)\right|}{\pi\Omega^{2}}d\Omega \label{EQ2}
\end{equation}
where $H_{0}=H(0)$ is the amplitude-gain of the medium at the frequency
of the optical carrier. This confirms that large group delays are
achieved when the gain $\left|H(\Omega)\right|$ has a pronounced
maximum at $\Omega=0$. We have then $\kappa_{2}>0$. 

Coming back to the general problem, we characterise the time function $x(t)$ by its area 
$S=\int_{-\infty}^{\infty}x(t)dt$ and its three lowest order moments, namely the mean value 
$\left\langle t\right\rangle =\frac{1}{S}\int_{-\infty}^{\infty}t\: x(t)dt$, the variance 
$\sigma^{2}=\frac{1}{S}\int_{-\infty}^{\infty}(t-\left\langle t\right\rangle )^{2}x(t)dt$ 
and the $3^{rd}$ order centred moment $a=\frac{1}{S}\int_{-\infty}^{\infty}(t-\left\langle t\right\rangle )^{3}x(t)dt$ . 
We recognise in $\left\langle t\right\rangle$ (resp. $\sigma$) the location of the centre-of-gravity 
(resp. the \emph{rms} duration) of the function $x(t$). Its asymmetry may be characterised by the dimensionless 
parameter $\xi=a/\sigma^{3}$, the so-called skewness \cite{ref20}. For a Gaussian function, $\sigma=\tau/(2\sqrt{\ln2})$ 
where $\tau $ is the FWHM of the energy profile $x^{2}(t)$. An important result \cite{ref20} is that the 
moments $\left\langle t\right\rangle$ , $\sigma^{2}$ and $a$ of $x(t)$ are equal to the cumulants $\kappa_{1},\kappa_{2}$ and 
$\kappa_{3}$ of $X(\Omega)$. Moreover the transfer equation immediately leads to the relations $E_{out}(0)=H_{0}E_{in}(0)$ and 
$\kappa_{n,out}=\kappa_{n,in}+\kappa_{n}$ where, as in all our paper, the indexes $in$, $out$ and the absence of index 
respectively refer to the incident pulse, the transmitted pulse and the transfer-function or impulse response of the medium. 
With our choice of time origin, $\left\langle t_{in}\right\rangle =0$. By combining the previous results, we finally obtain the four equations 
$S_{out}=H_{0}S_{in}$, $\left\langle t_{out}\right\rangle =\tau_{g}$, $\sigma_{out}^{2}=\sigma_{in}^{2}+\sigma^{2}>\sigma_{in}^{2}$
 and $a_{out}=a_{in}+a$. In the studies of the linear pulse propagation, the first equation which relates the areas of the transmitted 
and incident pulses is known as the area theorem \cite{ref23}. The second equation expresses that the time-delay of the pulse 
centre-of-gravity equals the group delay \cite{ref22}. The two last ones specify how the \emph{rms} duration and the asymmetry of the 
incident pulse are modified by the medium. All these results are valid provided that the involved moments are finite \cite{ref19}.

\section{ANALYTIC EXPRESSIONS OF THE MEDIUM IMPULSE-RESPONSE}

In order to obtain a complete information on the shape and the amplitude
of the transmitted pulse, we obviously have to specify the complex
gain-factor $\Gamma(\Omega)$ of the medium. For the medium with a
resonant gain, it reads
\begin{equation}
 \Gamma(\Omega)=p_{N}(\Omega)G/2-A/2 \label{EQ3} 
\end{equation} 
where $p_{N}(\Omega)$ is the normalised complex profile of the gain-line ($p_{N}(0)=1$), $G$
is the gain parameter for the intensity \cite{ref14} and $A$ stands for
the attenuation introduced to reduce the effects of the amplified
spontaneous emission \cite{ref15} and/or to normalise the overall gain
of the system. $G=g_{0}L$ where $g_{0}$(resp.$L$) is the resonance
gain-coefficient (resp. the thickness) of the medium. The previous
expression of $\Gamma(\Omega)$ also holds for an absorbing medium
with a transparency-window when the absorption background is assumed
to be infinitely wide. We then get $\Gamma(\Omega)=-[1-f\: p_{N}(\Omega)]\alpha_{0}L/2$
where $\alpha_{0}$ is the background absorption-coefficient, $f\leq1$
specifies the depth of the transparency-window \cite{ref9} and $p_{N}(\Omega)$
is the normalised complex profile of the line associated with the transparency-window.
By putting $G=f\:\alpha_{0}L$ and $A=\alpha_{0}L$ , we actually
retrieve the expression of $\Gamma(\Omega)$ obtained for a gain medium.
In both types of experiments, $G$ and $A$ are generally comparable
in order that the resonance gain $H_{0}=\textrm{e}^{G/2-A/2}$ is close to
1 or, at least, does not differ too strongly from 1. Anyway the intensity-transmission
on resonance exceeds its value far from resonance by the factor $\textrm{e}^{G}$. 

To go beyond, it seems necessary to explicit the profile $p_{N}(\Omega)$.
We first consider the reference case where $p_{N}(\Omega)$ is associated
with a Lorentzian line. It then reads $p_{N}(\Omega)=1/(1+i\Omega/\gamma)$
where $\gamma$ is the half-width of the line \cite{ref22} and we immediately
get $\kappa_{n}=Gn!/(2\gamma^{n})$ with in particular $\kappa_{1}=\tau_{g}=G/(2\gamma)$,
$\kappa_{2}=\sigma^{2}=G/\gamma^{2}$ and thus $\tau_{g}=\sigma\sqrt{G}/2$.
The last relation shows that achieving substantial fractional delays
$\tau_{g}/\sigma$ requires that $G\gg1$. A quite remarkable property
of the Lorentzian case is that the impulse response has an exact analytical
expression. This result has been obtained by Crisp in a general study
on the propagation of small-area pulses in absorbing and amplifying
media \cite{ref23} but it can easily be retrieved from $H(\Omega)$ by
using standard procedures of Laplace transforms \cite{ref20}. One get
\begin{equation}
 h(t)=\textrm{e}^{-A/2}\delta(t)+\textrm{e}^{-A/2}\gamma G\:\frac{I_{1}(\sqrt{2G\gamma t})}{\sqrt{2G\gamma t}}\: \textrm{e}^{-\gamma t}U(t)\label{EQ4} 
\end{equation} 
where $\delta(t)$, $I_{1}(u)$ and $U(t)$ respectively designate
the Dirac function, the $1^{st}$ order modified Bessel function and
the unit step function. The $1^{st}$ term $h_{i}(t)$ in $h(t)$
results from the constant value $\textrm{e}^{-A/2}$ of $H(\Omega)$ far from
resonance. This part of the response is instantaneous in our local
time picture and only the $2^{nd}$ term $h_{d}(t)$, directly associated
with the transmission peak, contribute to the delay. The areas of
$h_{i}(t)$ and $h_{d}(t)$ are respectively $\textrm{e}^{-A/2}$ and $H_{0}-\textrm{e}^{-A/2}$,
that is in a very small ratio ($\approx \textrm{e}^{-G/2}$ ) for the large
values of $G$ required to achieve substantial delays (see above).
\begin{figure}[ht]
\begin{center}
    \includegraphics[angle=0,width=8cm]{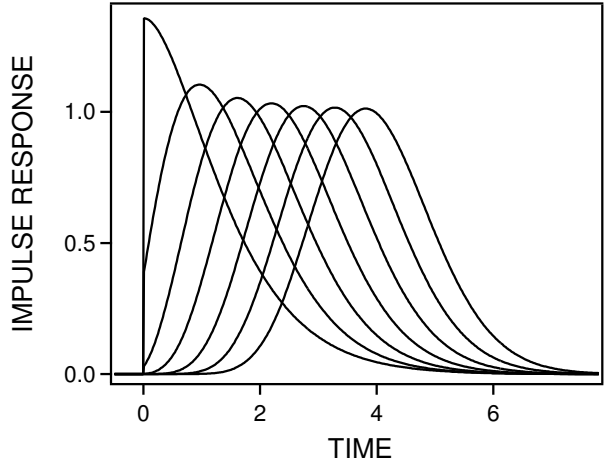}
    \caption{Analytical form of the impulse response for a Lorentzian line. From the left to the right the gain 
parameter $G$ (resp. the fractional delay $\tau_{g}/\sigma $) is 4, 9, 16, 25, 36, 49 and 64 (resp. 1, 1.5, 2, 2.5, 3, 3.5 and 4). 
The horizontal (resp. vertical) unit is $\sigma$  (resp. $H_{0}/\sigma \sqrt{2 \pi}$  ). \label{fig1}}
\end{center}
\end{figure}
The effect of the instantaneous response then becomes negligible.
Fig.\ref{fig1} shows the delayed response obtained for increasing values of
the gain and thus of the fractional group-delay. We see that the curves,
first strongly asymmetric, become more and more symmetric as $G$
increases and that the location of their maximum then approaches the
group delay. They have a discontinuity $H_{0}e^{-G/2}\gamma G/2$
at $t=0$, the relative amplitude of which becomes negligible when
$G\gg1$. From the asymptotic behaviour of $I_{1}(u)$ \cite{ref20}, we
then get 
\begin{equation}
 h_{d}(t)\approx \frac{H_{0}}{\sigma\sqrt{2\pi}}\left(1-\frac{3\theta}{4\tau_{g}}\right)\exp\left( -\frac{\theta^{2}}{2\sigma^{2}}\right)\label{EQ5} 
\end{equation}
with $\theta=t-\tau_{g}$. The maximum of $h_{d}(t)$ occurs at the instant
$\tau_{g}-\Delta t$ with $\Delta t\approx3\sigma^{2}/(4\tau_{g})$.
When $G\rightarrow\infty$, 
\begin{equation}
 h(t)\rightarrow h^{(2)}(t)=\frac{H_{0}}{\sigma\sqrt{2\pi}}\exp\left(-\frac{\theta^{2}}{2\sigma^{2}}\right).\label{EQ6} 
\end{equation}
This Gaussian form is that of the normal distribution derived by means of
the central limit theorem in probability theory. This theorem can
also be used for an approximate evaluation of the convolution of $n$
deterministic functions \cite{ref19}. It applies to our case by splitting
the medium in $n$ cascaded sections, $h(t)$ being then the convolution
of the impulse responses of each section. According to this analysis,
one may expect that the normal form $h^{(2)}(t)$ is universal. From
the frequency viewpoint, it originates in the fact that, when $G\gg1$
, the transmission peak is roughly $\sqrt{G}$ times narrower than
the line. In the region where the relative gain $\left|H(\Omega)\right|/H_{0}$
is not negligible, the curves $\Phi(\Omega)$ vs $\Omega$ (phase-shift)
and $F(\Omega)$ vs $\Omega$ (line-profile) are well approximated
respectively by a straight line and a parabola (Fig.\ref{fig2}). 
\begin{figure}[ht]
\begin{center}
    \includegraphics[angle=0,width=7cm]{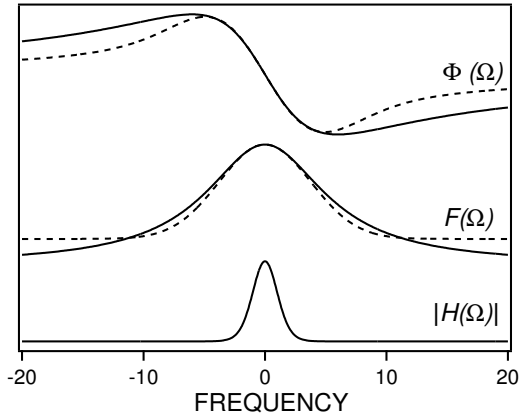}
    \caption{ $ \Phi( \Omega)$ and $F( \Omega)$ associated with a Lorentzian line (full line) and a Gaussian line (dashed line). 
The physical parameters are chosen in order that  $H_{0}=0.6$ and  $\tau_{g}/\sigma = 3$ in both cases. The resulting gain profiles 
$ \lvert H(\Omega) \rvert$  are indistinguishable at the figure scale. The frequency unit is $1/\sigma$ (angular frequency). \label{fig2}}
\end{center}
\end{figure}
This means that only the first two cumulants $\kappa_{1}=\tau_{g}$ and $\kappa_{2}=\sigma^{2}$
play a significant role. We then get $H(\Omega)\approx H^{(2)}(\Omega)=H_{0}\exp(-i\Omega\tau_{g}-\sigma^{2}\Omega^{2}/2)$
and thus $h(t)\approx h^{(2)}(t)$ irrespective of the line-profile.
This confirms the universality of the normal form of $h(t)$ when
$G\rightarrow\infty$. In fact, $h^{(2)}(t)$ is a good approximation
of the exact impulse-response for the gains currently achieved in
the experiments. Fig.\ref{fig3} shows the result obtained with Lorentzian and
Gaussian line-profiles when $\tau_{g}=3\sigma$. In the second case,
$p_{N}(\Omega)=\exp(-\Omega^{2}/\gamma^{2})-2iD(\Omega/\gamma)/\sqrt{\pi}$
with $D(u)=\textrm{e}^{-u^{2}}\int_{0}^{u}\textrm{e}^{v^{2}}dv$ \cite{ref24} and the first
cumulants read $\kappa_{1}=G/(\gamma\sqrt{\pi})$, $\kappa_{2}=G/\gamma^{2}$
and $\kappa_{3}=4G/(\gamma^{3}\sqrt{\pi})$. The parameters
A, G and $\gamma$ are chosen such that $H_{0}$, $\tau_{g}$ and
$\sigma$ have the same values in both cases. Though the line-profiles
are quite different (see Fig.\ref{fig2}), the impulse responses are both close
to the normal form $h^{(2)}(t)$. Similar results (not shown) are
obtained with other line-profiles, including the EIT profile (see
hereafter). 

When the gain parameter is large but not very large, a better approximation
of the impulse response is obtained by considering the effect of the
$3^{rd}$ cumulant $\kappa_{3}$, equal to the asymmetry parameter
$a$ of $h(t)$. Provided that this effect may be considered as a
small perturbation, $H(\Omega)\approx H^{(3)}(\Omega)\approx\left(1+i\kappa_{3}\Omega^{3}/3!\right)H^{(2)}(\Omega)$.
From the correspondence $i\Omega\leftrightarrow d/dt$, we finally get
\begin{equation}
h(t)\approx h^{(3)}(t)\approx\left(1-\frac{a\theta}{2\sigma^{4}}\right)h^{(2)}(t). \label{EQ7} 
\end{equation}
 This result generalises that obtained in the Lorentzian case where
$a=3G/\gamma^{3}$ and $a/(2\sigma^{4})=3/(4\tau_{g})$.
In the Gaussian case, we find $a/(2\sigma^{4})=2/(\pi\tau_{g})$,
a value not far from the previous one. This explains why the two impulse
responses are very close (see Fig.\ref{fig3}). Anyway they are very well approximated
by $h^{(3)}(t)$ in each case. Quite generally the maximum of $h^{(3)}(t)$
occurs at $\tau_{g}-\Delta t$ with $\Delta t\approx a/(2\sigma^{2})=\xi\sigma/2$
where $\xi$ is the skewness of $h(t)$. In all the above calculations
we have implicitly assumed that $\Delta t$ is small compared to $\sigma$.
This implies that $\left|\xi\right|\ll2$ but we checked that $h^{(3)}(t)$
keeps a fairly good approximation of $h(t)$ for skewness up to $1$. 
\begin{figure}[ht]
\begin{center}
    \includegraphics[angle=0,width=8cm]{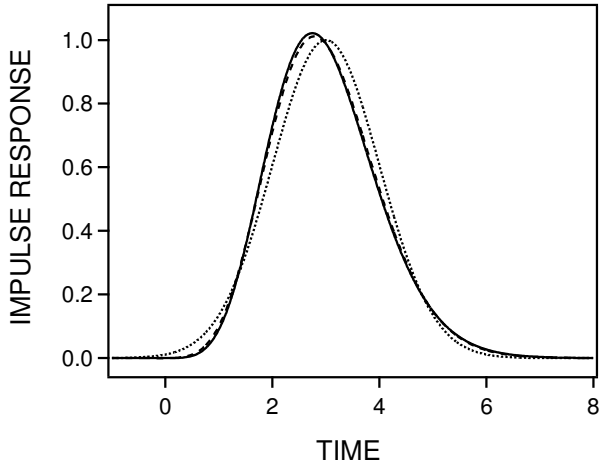}
    \caption{Comparison of the exact impulse-responses for the Lorentzian profile (full line) and the Gaussian profile 
(dashed line) to the normal form  $h^{(2)}(t)$ (dotted line). The fit by the improved forms  $h^{(3)}(t)$ (not shown for clarity)
 is nearly perfect. Parameters as in Fig.\ref{fig2}. Units as in Fig.\ref{fig1}. \label{fig3}}
\end{center}
\end{figure}

\section{NORMAL FORM OF THE TRANSMITTED PULSE} 
The impulse response being known, the envelope of the transmitted
pulse is given by the relation $e_{out}(t)=h(t)\otimes e_{in}(t)$
and will generally differ from that of the incident pulse. However
the distortion will be negligible if the duration of $h(t)$ is small
compared to that of the pulse. We then get $h(t)\approx\delta(t-\tau_{g})\int_{-\infty}^{+\infty}h(t)dt$
and thus $e_{out}(t)\approx H_{0}\: e_{in}(t-\tau_{g})$. Since we
are interested in the situations where the time delay is large compared
to the pulse duration, we obtain the double condition $\sigma\ll\sigma_{in}\ll\tau_{g}$
which can only be met with extremely large gain parameters. Taking
for example $\sigma=\sigma_{in}/7$ and $\tau_{g}=7\sigma_{in}$ that
is $\tau_{g}=49\sigma$, we get $G\approx9600$ in the Lorentzian
case. Fig.4 shows the results obtained with these parameters. As expected
the pulse distortion is small, even in the sensitive case of a square-shaped
pulse. Note that the gain parameter considered is not unrealistic.
It is comparable to that used by Harris and co-workers in their pioneering
EIT experiment where $G\approx A\approx6000$ \cite{ref2}. 
\begin{figure}[ht]
 \begin{center}
   \includegraphics[angle=0,width=8cm]{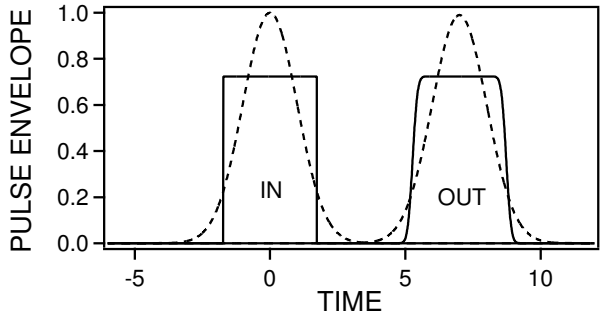}
   \caption{Propagation of a square-shaped (full line) and a Gaussian-shaped (dashed line) light-pulse with large delay 
and low distortion. The parameters are chosen in order that  $\sigma = \sigma_{in}/7$, $\tau_{g} = 7 \sigma_{in}$ 
and $H_{0} = 1$. The envelopes of the input pulses are given for reference. The time unit is the common \emph{rms} 
duration $\sigma_{in}$ of the two incident pulses. As expected, the distortion of the Gaussian-shaped pulse 
is negligible and there is only a slight softening of the rise and of the fall of the square-shaped pulse. \label{fig4}}
\end{center}
\end{figure}

\begin{figure}[ht]
 \begin{center}
   \includegraphics[angle=0,width=8cm]{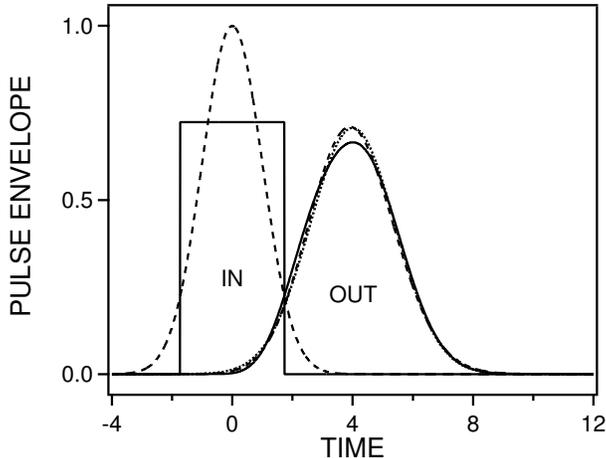}
   \caption{Example of pulse reshaping and broadening. The square-shaped (full line) and Gaussian-shaped (dashed line) 
incident pulses originate nearly identical transmitted pulses, respectively close and very close to the normal form 
$e_{out}^{(2)}(t)$ (dotted line). The time unit is $\sigma_{in}$  and the parameters are such that $\sigma = \sigma_{in}$, 
$\tau_{g}/\sigma = 4 $ and $H_{0} = 1$. With this choice of $H_{0}$ and $\sigma$, the transmitted pulses have the 
same area that the incident ones (area theorem) and a \emph{rms} duration $\sqrt{2}$ times larger.\label{fig5}}
 \end{center}
\end{figure}

However most of the direct demonstrations of large fractional pulse
delays have been achieved with smaller gain parameters, typically
ranging from 10 to 100, and with incident pulses whose \emph{rms}
duration $\sigma_{in}$ is comparable to and often smaller than $\sigma$.
Substantial pulse-reshaping is then expected. Suppose first that the
normal form $h^{(2)}(t)$ provides a good approximation of the impulse
response and that the incident pulse is Gaussian-shaped. We have then
$e_{in}(t)=\exp(-t^{2}/2\sigma^{2})$ and $e_{out}(t)$, convolution
of two Gaussian functions, is itself Gaussian. 
It reads $e_{out}(t)=H_{0}(\sigma_{in}/\sigma_{out})\exp(-\theta^{2}/2\sigma_{out}^{2})$,
where $\sigma_{out}^{2}=\sigma_{in}^{2}+\sigma^{2}$ . The effect
of the medium on the light-pulse is simply to delay its maximum exactly
by the group delay, to broaden it by the factor $\sqrt{1+\sigma^{2}/\sigma_{in}^{2}}$
{[}1, 9{]} and to modify its amplitude accordingly in order to respect
the area theorem \cite{ref23}. Since this point is often overlooked, we
stress that the broadening mechanism radically differs from that occurring
in standard optical fibres \cite{ref25}. It originates in the $2^{nd}$
order gain-dispersion instead of in the group-velocity dispersion
and the pulse envelope keeps real (no phase modulation or frequency
chirping). In fact, provided that $\sigma_{in}$ be smaller than or
comparable to $\sigma$ and that $\left|\xi_{in}\right|<1$, $e_{out}(t)$
is well approximated by a Gaussian function whatever the shape of
the incident pulse is. This is again a consequence of the central
limit theorem, the response $e_{out}(t)$ being obtained by an extra
convolution added to those used to build $h(t)$. The conditions on
$\sigma_{in}$ and $\left|\xi_{in}\right|$ originate in the requirement
that all the terms to convolute should have moments of the same order
of magnitude. We then obtain $e_{out}(t)\approx e_{out}^{(2)}(t)$
where $e_{out}^{(2)}(t)$ has the normal (Gaussian) form 
\begin{equation}
e_{out}^{(2)}(t)=H_{0}\frac{S_{in}}{\sigma_{out}\sqrt{2\pi}}\exp\left(- \frac{\theta^{2}}{2\sigma_{out}^{2}}\right). \label{EQ8} 
\end{equation}
This result extends the previous one and shows that incident pulses
having different shapes but the same area $S_{in}$ and the same variance
$\sigma_{in}$ are reshaped in the medium to give approximately the
same Gaussian-shaped pulse (Fig.\ref{fig5}). From an experimental viewpoint,
the dramatic reshaping of a square-shaped pulse has been clearly demonstrated
(but not commented on) by Turukhin \emph{et al}. \cite{ref8} in their EIT experiment
in a solid (see their figure 2c for 0 probe detuning). Pulse reshaping
is also apparent in the Brillouin scattering experiment by Song \emph{et
al}. \cite{ref13} where a flat-topped pulse is actually transformed in
a gaussian-like pulse (see their figure 4 and compare the shapes obtained
for gains 0dB and 30dB). 

\begin{figure}[ht]
 \begin{center}
   \includegraphics[angle=0,width=8cm]{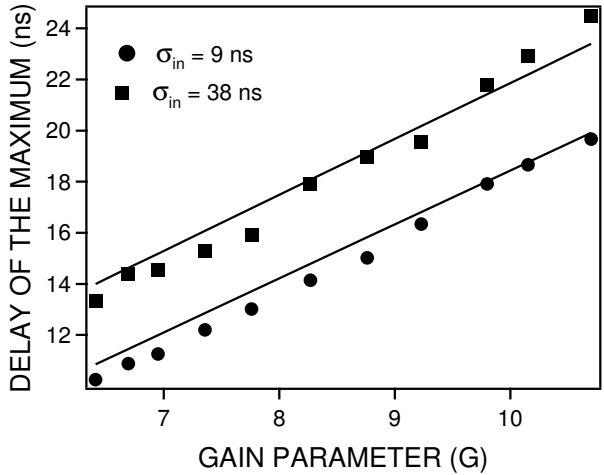}
   \caption{Comparison of the delays of the maximum of the transmitted pulse observed in a Brillouin scattering 
experiment \cite{ref14} (filled squares and circles) with our analytical predictions (full lines).\label{fig6}}
\end{center}
\end{figure}

A more precise approximation $e_{out}^{(3)}(t)$ of $e_{out}(t)$
can be obtained by taking into account the effect of the $3^{rd}$
order cumulants. Using the approach already used to determine $h^{(3)}(t)$,
we get
\begin{equation}
 e_{out}^{(3)}(t)\approx \left(1-\frac{a_{out} \theta}{2 \sigma_{out}^{4}}\right ) e_{out}^{(2)}(t) \label{EQ9} 
\end{equation}
with $a_{out}=a_{in} + a$. When the incident pulse is symmetric ($a_{in}=0$
) as in most experiments, the skewness $\xi_{out}$ of the transmitted
pulse reads $\xi_{out}=a/\left(\sigma^{2}+\sigma_{in}^{2}\right)^{3/2}$
and the pulse maximum occurs at $\tau_{g}-\Delta t_{out}$ with $\Delta t_{out}=a/\left[2\left(\sigma^{2}+\sigma_{in}^{2}\right)\right]$.
Since $\left|\xi_{out}\right|<\left|\xi\right|$, the transmitted
pulse is closer to a normal form than the impulse response of the
medium. The previous results hold without restriction to the value
of $\sigma_{in}$ when $e_{in}(t)$ is Gaussian. In the case of a
Lorentzian line-profile, we easily get 
\begin{equation}
\Delta t_{out}=\frac{3}{2\gamma\left(1+\sigma_{in}^{2}\gamma^{2}/G\right)}. \label{EQ10} 
\end{equation}
We have compared the theoretical delay of the pulse maximum, namely
$\tau_{g}- \Delta t_{out}$, with the delay actually observed by Okawachi
\emph{et al.} in their Brillouin scattering experiment \cite{ref14}. Fig.\ref{fig6}
shows this delay as a function of the gain parameter $G$ for two
values of the pulse duration, respectively $\tau_{in} = 63\: ns$ ($\sigma_{in}\approx 38\: ns$) 
and $\tau_{in}=15\: ns$ ($\sigma_{in}\approx 9\: ns$ ), with $\gamma = 0.22\: ns^{-1}$
($\gamma$ is the half of the full Brillouin linewidth $\Gamma_{B}$).
Without any adjustment of parameters, our analytical results satisfactorily
fit the observations. Note that the shifts $\Delta t_{out}$ are negligible
for the longer pulse but significant for the shorter one. 

\section{EFFECT OF THE TRANSMISSION BACKGROUND} 
 
In the previous calculations, we have not taken into account the effect
of the instantaneous part $h_{i}(t)=\textrm{e}^{-A/2}\delta(t)$ of the impulse
response, arguing that its area is small compared to that of the delayed
part. As a matter of fact, $h_{i}(t)$ originates a contribution $\textrm{e}^{-A/2}e_{in}(t)$
to $e_{out}(t)$, the amplitude of which is roughly $\textrm{e}^{G/2}\sigma_{in}/\sigma_{out}$
times smaller than that of the main part and is thus actually negligible
in every case of substantial delay ($G\gg1$, $\sigma_{out}$ and
$\sigma_{in}$ of the same order of magnitude). We should however remark
that this result lies on the assumption that the transmission peak
is put on an uniform background.

We will now examine the case where the transmission background is
not uniform. This happens in all the experiments where a transparency-window
is induced in a absorption-profile of finite width, in particular
in the EIT experiments. As an illustrative example we consider the
simplest $\Lambda$ arrangement with a resonant control field. From
the results given in \cite{ref4}, we easily get
\begin{equation}
\Gamma(\Omega)=-\frac{\gamma_{ba}\left(i\Omega+\gamma_{ca}\right)A/2}{\left(i\Omega+\gamma_{ba}\right)\left(i\Omega+\gamma_{ca}\right)+\Omega_{s}^{2}/4} \label{EQ11} 
\end{equation} 
where $\gamma_{ba}$ (resp. $\gamma_{ca}\ll\gamma_{ba}$) is the coherence
relaxation-rate for the probe transition (resp. for the forbidden
transition), $\Omega_{s}$ is the modulus of the Rabi frequency associated
with the control field and $A=\alpha_{0}L\gg1$ is the resonance optical
thickness in the absence of control field. The control field makes
the resonance gain rise from $\textrm{e}^{-A/2}\approx0$ to 
$H_{0}=\exp\left(-A\gamma_{ba}\gamma_{ca}/\left[2\left(\gamma_{ba}\gamma_{ca}+\Omega_{s}^{2}/4\right)\right]\right)$
and a good transparency is induced when $\Omega_{s}$ is larger than
or comparable to $\gamma_{ba}\gamma_{ca}A$. The width of the transparency-window
($\propto$ $\Omega_{s}$/$\sqrt{A}$ ) is then much smaller than that
of the absorption background ($\propto\gamma_{ba}\sqrt{A}$). 

\begin{figure}[ht]
 \begin{center}
   \includegraphics[angle=0,width=8cm]{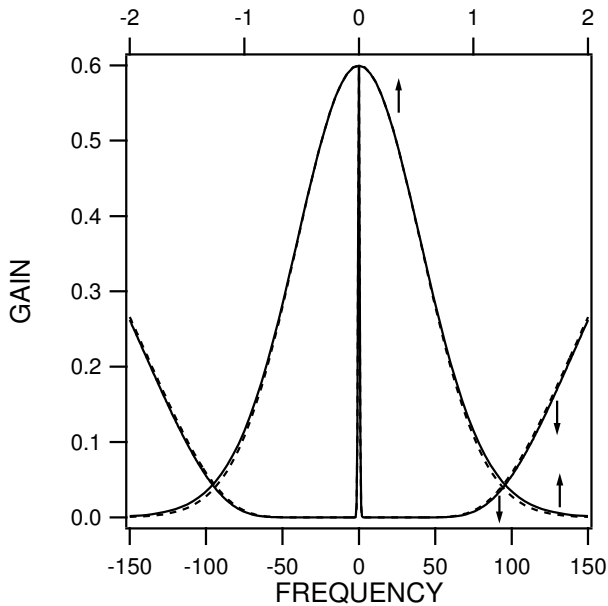}
   \caption{Gain profile in an EIT experiment and expanded view of its central part (upper scale). In the wings 
as in the transparency window, the exact gain $ \lvert H(\Omega) \rvert$ (full line) is scarcely distinguishable 
from the approximate form $ \lvert H^{(2)}(\Omega)+ H_{off}(\Omega)\rvert$ (dashed line). 
Remind that $ \lvert H(\Omega) \rvert \rightarrow 1$ in the far wings. The parameters are $A=63$, $\gamma_{ba}/2\pi=5$ MHz, 
$\gamma_{ca}/\gamma_{ba}= 2.2 \: 10^{-3}$ and $\Omega_{s}/\gamma_{ba}= 0.73$ . The frequency unit is $10^{6}$ Rd/s
(angular frequency).\label{fig7}}
\end{center}
\end{figure}

Without any approximation, the partial fraction decomposition of $\Gamma(\Omega)$
allows us to write the transfer-function of the medium as a product
of simpler functions, namely $H(\Omega)=H_{1}(\Omega)H_{2}(\Omega)$
with $H_{j}(\Omega)=\exp\left[C_{j}/\left(i\Omega+\gamma_{j}\right)\right]$.
According to the control power, the parameters $C_{1}$, $C_{2}$,
$\gamma_{1}$ and $\gamma_{2}$ are real or complex. When $\Omega_{s}<(\gamma_{ba}-\gamma_{ca})$,
$\gamma_{1}$ and $\gamma_{2}$ are real and positive whereas $C_{1}$
and $C_{2}$ are also real but of opposite sign. The EIT medium is
then equivalent to a medium with two Lorentzian lines both centred
at $\Omega=0$, respectively an absorption-line and a narrower gain-line.
It is also equivalent to a cascade of a gain medium and an absorbing
medium. When $\Omega_{s}>(\gamma_{ba}-\gamma_{ca})$, all the parameters
are complex with $\gamma_{2}=\gamma_{1}^{*}$ and $C_{2}=C_{1}^{*}$
. The two lines are now located at $\Omega=\pm\textrm{Im}(\gamma_{1})$.
They have the same intensity and the same width, but they are hybrid
in the sense that, due to the complex nature of $C_{1}$ and $C_{2}$,
their absorption and dispersion profiles are both the sum of an absorption-like
and a dispersion-like profile. We incidentally note that the parameters
used to obtain the figure 8 in \cite{ref4} correspond to such a situation.
In all cases, the impulse responses associated to $H_{1}(\Omega)$
and $H_{2}(\Omega)$ have analytical expressions \cite{ref23,ref26} and
the impulse response $h(t)$ of the medium is their convolution product.
This general analysis is satisfactory from a formal viewpoint. It
provides some physical insight into the EIT mechanisms but is not
really operational to determine the shape of the transmitted pulse.
From this viewpoint, a fruitful approach consists in exploiting the
fact that the medium is opaque except in the narrow region of induced
transparency and in the far wings of the background absorption-line
(Fig.\ref{fig7}). 
In the first region, $H(\Omega)$ is well approximated by
the forms $H^{(2)}(\Omega)$ or, if necessary, $H^{(3)}(\Omega)$
, obtained by keeping only the 2 or 3 first cumulants of $H(\Omega)$
as in the case of an uniform background. We only give here the simplified
expressions of these cumulants when $\gamma_{ca}\ll\gamma_{ba}$ and 
$\Omega_{s}^{2}\gg\gamma_{ca}\gamma_{ba}$ (conditions of good induced
transparency). We then get $\kappa_{1}=\tau_{g}\approx2A\gamma_{ba}$/$\Omega_{s}^{2}$,
$\kappa_{2}=\sigma^{2}\approx16A\gamma_{ba}^{2}$/$\Omega_{s}^{4}$
and $\kappa_{3}=a\approx48A\gamma_{ba}(4\gamma_{ba}^{2}-\Omega_{s}^{2})$/$\Omega_{s}^{6}$
with $H_{0}\approx\exp\left(-2A\gamma_{ba}\gamma_{ca}/\Omega_{s}^{2}\right)$.
In the far wings $\left|\Omega\right|\gg\Omega_{s}$ and $H(\Omega)\approx H_{off}(\Omega)$,
where $H_{off}(\Omega)=\exp\left(-A/[2(1+i\Omega/\gamma_{ba})]\right)$
is the transfer-function when the control field is off. Finally we
get the relation $H(\Omega)\approx H^{(p)}(\Omega)+H_{off}(\Omega)$
with $p=2$ or $3$, valid at every frequency. Fig.\ref{fig7}, obtained for
typical physical parameters, shows that $\left|H^{(2)}(\Omega)+H_{off}(\Omega)\right|$
already provides a good approximation of the exact gain. Now reduced
to a simple sum instead of a convolution product, the medium impulse-response
reads $h(t)=h^{(p)}(t)+h_{off}(t)$ where $h_{off}(t)$, associated
with a Lorentzian absorption-line, has an analytical expression \cite{ref23}.
As in the case of a gain-line, this expression can be retrieved from
$H_{off}(\Omega)$ by using standard procedures of Laplace transforms
\cite{ref20}. It reads
\begin{equation}
h_{off}(t)=\delta(t)-\gamma_{ba}A\frac{J_{1}(\sqrt{2A\gamma_{ba}t})}{\sqrt{2A\gamma_{ba}t}}\exp\left(-\gamma_{ba}t\right)U(t)\label{EQ12} 
\end{equation}  
where $J_{1}(u)$ designates the ordinary $1^{st}$ order Bessel function.
The envelope $e_{out}(t)$ of the transmitted pulse will be thus the
sum of two terms. The first one is the approximate solution $e_{out}^{(p)}(t)$
obtained by the cumulants procedure. The second one reads $e_{off}(t)=h_{off}(t)\otimes e_{in}(t)$.
It is worth remarking that $h_{off}(t)$ is rapidly oscillating (characteristic
time $\propto1/A\gamma_{ba}$) and that its area is very small ($\int_{-\infty}^{+\infty}h_{off}(t)dt=H_{off}(0)=\textrm{e}^{-A/2}$
). This entails that $e_{off}(t)$ will have a negligible amplitude
($\propto e^{-A/2}$) when $e_{in}(t)$ is smooth enough so that the
far wings of its Fourier spectrum do not overlap those of the absorption-line.
\begin{figure}[ht]
 \begin{center}
   \includegraphics[angle=0,width=8cm]{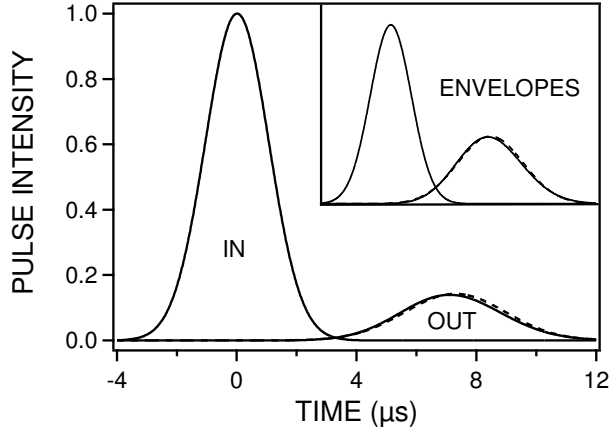}
   \caption{Propagation of a Gaussian-shaped pulse in the EIT experiment. The parameters are as in Fig.\ref{fig7} with
$ \sigma_{in}=1.5 \: \mu s $. The full and dashed lines respectively are the exact intensity-profile of the pulse and the 
normal form. The location of the two maximums differ by  $ 0.2 \: \mu s $ in agreement with 
the relation $ \Delta t_{out}=a/2 \sigma_{out}^{2} $. Inset : the corresponding pulse-envelopes 
for $-5 \: \mu s \leq t \leq 15 \: \mu s$ .\label{fig8}}
\end{center}
\end{figure}
Fig.\ref{fig8} shows the result obtained with a Gaussian-shaped incident-pulse. $A$
, $\gamma_{ba}$ and $\sigma_{in}$ being given, we have chosen the
other parameters in order to reproduce the location and the amplitude
of the maximum of the transmitted pulse in the celebrated experiment
by Hau \emph{et al.} \cite{ref7}. We then get $\sigma_{out}/\sigma_{in}\approx1.6$,
a broadening consistent with the observations, and $\xi_{out}\approx0.16$.
The asymmetry being very slight, the delay of the maximum is very
close to $\tau_{g}$ and $e_{out}(t)$ is well fitted by the normal
(Gaussian) form $e_{out}^{(2)}(t)$. A perfect fit is obtained by
using the improved form $e_{out}^{(3)}(t)$. The contribution $e_{off}(t)$
associated with $h_{off}(t)$ is actually too small to be visible.
Conversely $h_{off}(t)$ will be responsible for the generation of
short transients when $e_{in}(t)$ comprises localised defects. As
expected and recently discussed about the EIT experiments \cite{ref27},
the front of these transients will propagate at the velocity $c$
(instantaneously in our local time picture). Their peak amplitude
will be especially large when the defects consist in discontinuities.
Consider again a square-shaped incident-pulse, the total duration
$2\tau_{p}$ and the amplitude $\eta$ of which are such that its area and its
variance equal those of the Gaussian-shaped pulse (Fig.\ref{fig9}).
\begin{figure}[ht]
 \begin{center}
   \includegraphics[angle=0,width=8cm]{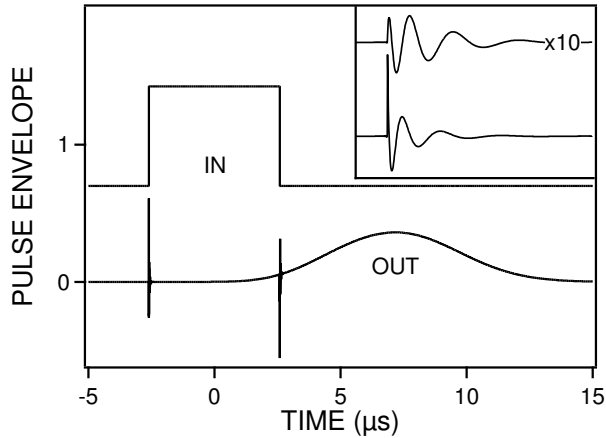}
   \caption{Propagation of a square-shaped pulse in the EIT experiment. Parameters as in the two previous figures. The upper 
and lower curves are the envelopes of the incident and transmitted pulses. The peak amplitude of the transients is slightly 
smaller than its theoretical value due to the finite time-resolution of the computations (0.4 ns). Inset : the first transient 
expanded on a $0.2 \mu s$ time-interval (bottom) and the same  ($\times 10$) after passage through a $ 1^{st}$ order filter of 
time-constant $ \sigma_{out}/100$ (top). Such a filtering does not significantly affect the smooth part of the pulse which keeps very 
close to that obtained with an incident Gaussian-shaped pulse but reduces the amplitude (resp. the intensity) of the transient by a 
factor of about 30 (resp. 900). \label{fig9}}
\end{center}
\end{figure} 
$e_{off}(t)$ is then easily derived from $h_{off}(t)$. It reads
\begin{equation}
  e_{off}(t)=\eta\left[f(t+\tau_{p})-f(t-\tau_{p})\right]\label{EQ13} 
\end{equation} 
with 
\begin{equation}
f(t')=U(t')-\gamma_{ba}A\int_{0}^{t'}\frac{J_{1}(\sqrt{2A\gamma_{ba}x})}{\sqrt{2A\gamma_{ba}x}}\exp\left(-\gamma_{ba}x\right)dx. \label{EQ14} 
\end{equation} 
Each discontinuity in $e_{in}(t)$ actually originates a large transient.
Its initial amplitude is equal to that of the incident pulse and its
successive maximums of intensity occur at the instants $j_{1n}^{2}/2A\gamma_{ba}$
later, $j_{1n}$ being the $n^{th}$ zero of $J_{1}(u)$ \cite{ref28}.
Note that the amplitude of the transients exceeds that of the smooth
part of $e_{out}(t)$. In a real experiment however the finite values
of the rise and fall times of the incident pulse and of the detection
bandwidth will generally limit the importance of the transients. By
a deliberate reduction of the detection bandwidth, it is even possible
to bring their intensity to a very low level without significantly
affecting the delayed Gaussian-like part (see inset of Fig.\ref{fig9}). 

\begin{figure}[ht]
 \begin{center}
   \includegraphics[angle=0,width=8cm]{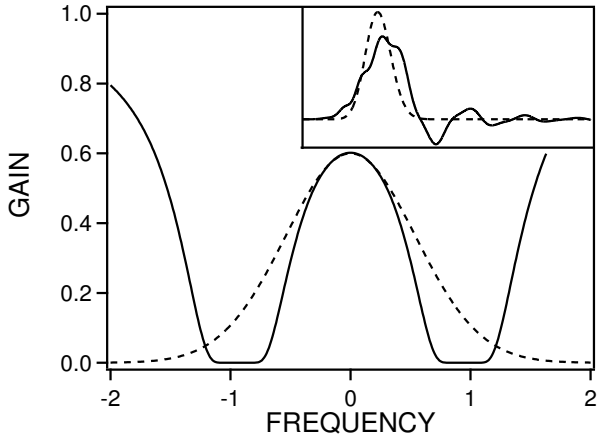}
   \caption{. Gain-profile $ \lvert H(\Omega) \rvert$ for the two absorption-lines arrangement (full line) compared to the 
corresponding normal gain-profile $ \lvert H^{2(}(\Omega) \rvert$  (dotted line). $H_{0}$, $\tau_{g}$, $\sigma$  and frequency 
unit as in Fig.\ref{fig7}. Inset : envelope of the transmitted pulse for a Gaussian-shaped incident pulse (full line) compared 
to that obtained with the EIT-arrangement (dotted line). $\sigma_{in}=1.5\:\mu s$   \label{fig10}}
\end{center}
\end{figure}

The results obtained on the model EIT-arrangement hold for an extended
class of systems having a transparency-window in a wide absorption
profile. They only lie on three assumptions: (i) $\Gamma(\Omega)$
is Lorentzian in the far wings of the absorption profile (ii) the
opaque regions are much wider that the transparency-window (iii) the
transfer-function does not significantly deviate from the normal form
in the transparency-window. The first condition (i) is generally met
even when $\Gamma(\Omega)$ is not Lorentzian in its central part.
Anyway, it is not essential. If it is not met, the detailed shape
of the transients is modified but not their main features (instantaneous
transmission, duration proportional to the inverse of the spectral
width of the opaque regions). The conditions (ii) and (iii), which
are closely related, are met in the EIT experiments when the medium
transmission is good at the frequency of the optical carrier (see
before) but this is not always sufficient. As a counter-example we
consider the experiment achieved by Tanaka \emph{et al}. in an atomic vapour
with a natural transparency-window between two strong absorption lines
\cite{ref29}. The complex gain-factor reads
\begin{equation}
\Gamma(\Omega)=-\frac{A}{2}\left[\frac{1}{1+i(\Omega+\Delta)/\gamma}+\frac{1}{1+i(\Omega-\Delta)/\gamma}\right]\label{EQ15} 
\end{equation} 
where $2\Delta$ is the doublet splitting. Despite an apparent similarity,
the associated transfer-function dramatically differs from that obtained
in EIT when $\Omega_{s}>(\gamma_{ba}-\gamma_{ca})$. Indeed the two
involved lines are here purely Lorentzian (not hybrid) and a good
transparency at $\Omega=0$ is achieved only if $\Delta\gg\gamma$.
We then get $H_{0}\approx\exp\left(-A\gamma^{2}/\Delta^{2}\right)$,
$\tau_{g}\approx A\gamma/\Delta^{2}$ , $\sigma^{2}\approx6A\gamma^{2}/\Delta^{4}$,
$a\approx-6A\gamma/\Delta^{4}$ and $\xi\approx-\Delta^{2}/(\gamma^{2}\sqrt{6A})$.
Choosing the physical parameters such that $H_{0}$, $\tau_{g}$ and
$\sigma^{2}$ equal their values in the EIT experiment , we actually
obtain a quite different gain profile with opaque regions whose width
is smaller than that of the transparency-window (Fig.\ref{fig10}). 
This entails that the approximation $\left|H(\Omega)\right|\approx\left|H^{(2)}(\Omega)\right|$
only works in the immediate vicinity of $\Omega=0$. It is the same
for the law $\Phi(\Omega)\propto\Omega$, the skewness being very
large ($\xi\approx-7.6$ ). In such conditions, even a Gaussian-shaped
pulse is strongly distorted (see inset of Fig.\ref{fig10}). In fact the two
absorption-lines arrangement allows one to attain large fractional
delays $\tau_{g}/\sigma_{in}$ with moderate distortion and broadening
but this requires to involve much larger absorption-parameters and
to accept a lower transmission. For example, Tanaka \emph{et al.} succeeded
in obtaining $\tau_{g}/\sigma_{in}\approx13$ with Gaussian-like pulses
(see Fig.4c in \cite{ref29}) but the peak-intensity of the transmitted
pulse was 75 times smaller than that of the incident pulse. Their
results are well reproduced with our two-lines model by taking $A=2.6\times10^{4}$ and
$\Delta/\gamma=110$. We then get $\sigma^{2}/\sigma_{in}^{2}\approx1/25\ll1$
and we actually are in a case of low distortion as previously discussed
(see Fig.\ref{fig4}). However, pulses with discontinuities are excluded. Since
$H(\infty)\gg H_{0}$, the resulting transients (not delayed) would
indeed be much larger than the delayed part of the transmitted pulse
and would obscure it. Moreover, due to the narrowness of the opaque
regions, the time scales of the transients and of the delayed part
do not considerably differ and it is thus impossible to filter out
the former without denaturing the latter. 

\section{SUMMARY AND DISCUSSION} 
Privileging the time-domain analysis, we have studied the linear propagation
of light pulses, the frequency of which coincides with that of a pronounced
maximum in the transmission of the medium. An important point is that
substantial pulse-delays are only attained when the corresponding
transmission exceeds the minimum one by a very large factor $C$ (contrast).
The impulse response of the medium then tends to a normal (Gaussian)
form, irrespective of the line profile associated with the transmission
peak. The propagation of arbitrarily shaped light-pulses with significant
delays and low distortion is possible when the \emph{rms} duration
$\sigma$ of the medium impulse-response is small compared to that
of the incident pulse ($\sigma_{in}$), which should itself be small
compared to the group delay $\tau_{g}$. The fulfilment of this double
condition requires systems where $C$ is extremely large, typically
several tens of thousands of $dB$ in a logarithmic scale. Systems
with such contrast have actually been used \cite{ref2} but, in most slow-light
experiments, $C$ ranges from 30 to 600 dB. Significant fractional
time-delays $\tau_{g}/\sigma_{in}$ keep attainable with such values
of $C$ by using incident pulses, the duration of which is comparable
to or smaller than $\sigma$. As the medium impulse-response, the
transmitted pulse then tends to acquire a normal (Gaussian) shape
whatever its initial shape is. This reshaping is particularly striking
when the incident pulse is square-shaped but is reduced to a simple
broadening when the latter is itself Gaussian-shaped. Despite its
asymptotic character, the normal form generally provides a good approximation
of the shape of the transmitted pulse. More precise shapes are obtained
by a perturbation method, allowing us in particular to specify how
much the delay of the pulse-maximum deviates from the group delay.

All these results have been first established by assuming that the
transmission peak is put on an uniform background. We have shown that
they also apply when the transmission peak is associated with a transparency
window in an absorption-profile of finite width. This however requires
that the nearly opaque regions flanking the transparency window be
considerably wider than the latter. Other things being equal, there
are then no differences between the cases of uniform and non uniform
transmission-backgrounds, at least when the envelope of the incident
pulse is smooth. Conversely localised defects in this envelope will
be responsible for the generation of very short transients which complement
the normal (Gaussian) part of the signal. The front of the transients
is instantaneously propagated in our local time picture (that is at
the velocity $c$ in a dilute sample). In extreme cases, their amplitude
may be comparable to that of the delayed signal but, due to their
location and their duration, they can easily be eliminated without
altering the latter. 

The slow and fast light experiments have a common feature. In both
cases, the observation of significant effects requires media with a very large
contrast between the maximum and the minimum of transmission. This results from the 
causality principle and implies severe limits to the effects attainable in fast-light
experiments, whatever the involved system is \cite{ref30}. From this viewpoint the slow-light case is obviously less pathologic and the 
constraints, although real, are much softer. 

\section{ACKNOWLEDGEMENTS} 

Laboratoire PhLAM is Unit\'{e} Mixte de Recherche de l'Universit\'{e} de Lille I et du CNRS (UMR 8523).
 CERLA is F\'{e}d\'{e}ration de Recherche du CNRS
(FR 2416).

\end{document}